\documentclass[english,aps,prl,twocolumn,superscriptaddress,showpacs,reprint]{revtex4-1}
\usepackage{lmodern}
\usepackage[T1]{fontenc}
\usepackage[latin9]{inputenc}
\usepackage{color}
\usepackage{babel}
\usepackage{textcomp}
\usepackage{graphicx}
\usepackage{subscript}
\usepackage[unicode=true,pdfusetitle,
 bookmarks=true,bookmarksnumbered=false,bookmarksopen=false,
 breaklinks=false,pdfborder={0 0 0},backref=false,colorlinks=true]
 {hyperref}

\makeatletter

\@ifundefined{textcolor}{}
{%
 \definecolor{BLACK}{gray}{0}
 \definecolor{WHITE}{gray}{1}
 \definecolor{RED}{rgb}{1,0,0}
 \definecolor{GREEN}{rgb}{0,1,0}
 \definecolor{BLUE}{rgb}{0,0,1}
 \definecolor{CYAN}{cmyk}{1,0,0,0}
 \definecolor{MAGENTA}{cmyk}{0,1,0,0}
 \definecolor{YELLOW}{cmyk}{0,0,1,0}
}

\makeatother

\begin{document}

\title{Gapped Surface States in a Strong-Topological-Semimetal}

\author{A.P. Weber}

\email{apwnq5@mail.umkc.edu}

\affiliation{Department of Physics and Astronomy, University of Missouri-Kansas
City, Kansas City, Missouri 64110, USA}

\author{Q.D. Gibson}

\author{Huiwen Ji}

\affiliation{Department of Chemistry, Princeton University, Princeton, New Jersey
08544, USA}

\author{A.N. Caruso}

\affiliation{Department of Physics and Astronomy, University of Missouri-Kansas
City, Kansas City, Missouri 64110, USA}

\author{A.V. Fedorov}

\affiliation{Advanced Light Source, Lawrence Berkeley National Laboratory, Berkeley,
California 94720, USA}

\author{R.J. Cava}

\affiliation{Department of Chemistry, Princeton University, Princeton, New Jersey
08544, USA}

\author{T. Valla}

\affiliation{Condensed Matter Physics and Materials Science Department, Brookhaven
National Laboratory, Upton, New York 11973, USA}
\begin{abstract}
A three-dimensional strong-topological-insulator or -semimetal hosts
topological surface states which are often said to be gapless so long
as time-reversal symmetry is preserved. This narrative can be mistaken
when surface state degeneracies occur away from time-reversal-invariant
momenta. The mirror-invariance of the system then becomes essential
in protecting the existence of a surface Fermi surface. Here we show
that such a case exists in the strong-topological-semimetal Bi\textsubscript{4}Se\textsubscript{3}.
Angle-resolved photoemission spectroscopy and \textit{ab initio} calculations
reveal partial gapping of surface bands on the Bi\textsubscript{2}Se\textsubscript{3}-termination
of Bi\textsubscript{4}Se\textsubscript{3}(111), where an 85 meV
gap along $\bar{\Gamma}\bar{K}$ closes to zero toward the mirror-invariant
$\bar{\Gamma}\bar{M}$ azimuth. The gap opening is attributed to an
interband spin-orbit interaction that mixes states of opposite spin-helicity.
\end{abstract}


\maketitle
Topological insulator materials (TIM) \cite{Ando2013_Review}, which
include insulators \cite{Hasan2010(1),Hasan2011(2),Qi2011(3)} and
semimetals \cite{Hsieh(4),Hsieh(5),Sb(110)_STM_2012,Sb(110)ARPES,Bi4Se3S_2012,QDG2013},
possess an inverted bulk band gap that hosts unusually robust, spin-helical
topological surface states (TSS) at the material's boundaries. The
presence of TSS is guaranteed by the topology of the bulk bands and
the states can only be removed from the Fermi energy if protecting
symmetries are broken. Strong topological insulator (STI) \cite{Fu2007_WSTIS,FuInv}
materials hold a special distinction, as they are often said to host
gapless TSS protected by time-reversal symmetry (TRS) on every surface
termination of the crystal. We will show it is possible for the TSS
of material in a STI phase to intrinsically acquire a finite gap at
all momenta in the surface Brillouin zone (SBZ) which do not lie on
a mirror-invariant azimuth. In that case, the mirror-symmetry (MS)
of the crystal lattice must be intact to guarantee the existence of
a surface Fermi surface. Earlier work by Teo, Fu, and Kane \cite{TeoBiSb_2008}
had foreseen this possibility, however, no concrete examples have
been obtained in experiment. Here we show that such a system exists
in the strong-topological-semimetal Bi\textsubscript{4}Se\textsubscript{3}.
The finding of mirror-protected TSS in this system also provides direct
confirmation that the strong-topological-insulator phase can coexist
with the topological-crystalline-insulator phase of matter, as was
suggested by Rauch \textit{et al.} \cite{Rauch} for the case of Bi-chalcogenides.

Bi\textsubscript{4}Se\textsubscript{3} is a rhombohedral superlattice
material consisting of alternating Bi\textsubscript{2} and Bi\textsubscript{2}Se\textsubscript{3}
layers stacked along the (111) direction. Previously, Dirac cone-like
TSS were found in the gap between the first fully occupied bulk valence
band (BVB) and the hole-like bulk conduction band (HBCB) on two different
surface terminations of Bi\textsubscript{4}Se\textsubscript{3}(111)
\cite{QDG2013}. It was determined that the TSS result from a parity
inversion at the $\Gamma$-point of the bulk Brillouin zone (BBZ),
a characteristic shared with the Bi\textsubscript{2}Se\textsubscript{3}
parent compound \cite{SC zhang Bi2Se3 class}. Although effects of
hybridization between surface and bulk electrons were mentioned, it
was not made clear that the states of the upper Dirac cone appearing
on the Bi\textsubscript{2}Se\textsubscript{3}-terminated surface,
which have an electron-like dispersion, must have crossed the HBCB
to reach the Fermi level. The possibility of new topological constraints
on the surface electron-structure above the HBCB was also not explored.
Earlier work \cite{Bi4Se3S_2012} predicted that the band gap above
the HBCB is characterized by a single parity inversion at the F-point
of the BBZ, which demands that TSS within this gap must come in an
odd number of pairs \cite{TeoBiSb_2008}. If that is the case, then
an odd number of surface bands must appear within the gap to meet
the state which crossed the HBCB. Moreover, this parity inversion
is away from the center of the BBZ and the crystal does not cleave
at the center of inversion. Under these conditions, the TSS pairs
are not constrained to have Dirac points at time-reversal invariant
momenta (TRIM) \cite{TeoBiSb_2008}. 

Consistent with the prediction in ref. \cite{Bi4Se3S_2012}, it is
in the momentum-space region outside the HBCB edge that evidence of
partially gapped TSS is found through \textit{ab initio} calculations
and angle-resolved photoemission spectroscopy (ARPES) measurements
on the Bi\textsubscript{2}Se\textsubscript{3}-terminated surface
of Bi\textsubscript{4}Se\textsubscript{3}(111). The TSS degenerate
away from TRIM on the mirror-invariant $\bar{\Gamma}\bar{M}$ azimuth
in the SBZ. The bands become separated elsewhere, and an 85 meV gap
between the surface state branches is measured along $\bar{\Gamma}\bar{K}$
in ARPES. The origin of this gap is accounted for in a model for the
spin-orbit interaction on a (111) crystal surface. These findings
place 2D electron hybridization within the subject of pristine STI
materials and provide direct evidence for MS protection of surface
states in a Bi-chalcogenide.

Single crystals of Bi\textsubscript{4}Se\textsubscript{3} were synthesized
following a previously reported procedure \cite{QDG2013}. ARPES was
performed using a Scienta SES-100 electron spectrometer at beamline
12.0.1 of the Advanced Light Source with a combined instrumental energy
resolution of $\sim$12 meV and an angular resolution better than
$\pm$ 0.07\AA. The sample was cleaved under ultrahigh vacuum conditions
($<5.0\times10^{-9}$ Pa) and kept at $\sim$15 K. Temperature was
measured using a silicon sensor mounted near the sample.

Electron-structure calculations were performed in the framework of
density functional theory (DFT) using the WIEN2K code \cite{Wien2k_Ref}
with a full-potential linearized augmented plane-wave and local orbitals
basis together with the Perdew-Burke-Ernzerhof \cite{PBE_Ref} parametrization
of the generalized gradient approximation, using a slab geometry.
Experimentally determined lattice parameters and atom positions were
used to construct the slabs. The plane-wave cutoff parameter R\textsubscript{MT}K\textsubscript{max}
was set to 7 and the Brillouin zone was sampled by 9 \textit{k} points.
Spin-orbit coupling (SOC) was included. To study the Bi\textsubscript{2}Se\textsubscript{3}-terminated
surface, a slab was constructed of 6 Bi\textsubscript{2}Se\textsubscript{3}
layers and 5 Bi\textsubscript{2} layers, with 10 {\AA}  of vacuum between
adjacent slabs. The contribution of the surface atoms to the overall
surface electronic structure was determined by calculating the partial
contribution of each atomic basis set to the wave functions at all
\textit{k} points.

\begin{figure}
\includegraphics[width=80mm]{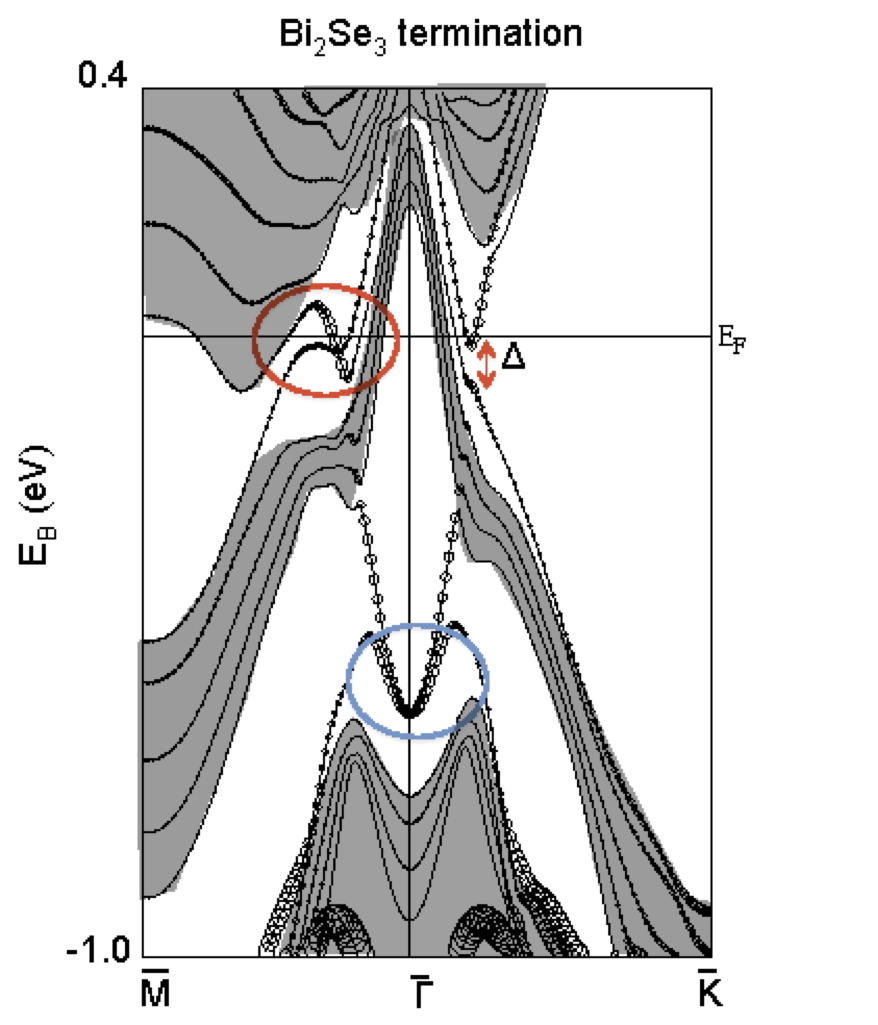}

\caption{Calculated band structure for the Bi\textsubscript{2}Se\textsubscript{3}-terminated
slabs of Bi\textsubscript{4}Se\textsubscript{3} plotted along the
$\bar{M}\bar{\Gamma}\bar{K}$ path in the surface Brillouin zone.
The size of the circular plotting markers indicate the contribution
from the surface layer. Shaded regions indicate the projection of
bulk electrons. The red circle contains the mirror-symmetry-protected
crossing of TSS and the red double arrow indicates the corresponding
TSS avoided crossing gap $\Delta$. The blue circle contains a
TSS crossing protected by both TRS and MS.}
\end{figure}

The calculated electron-structure for Bi\textsubscript{2}Se\textsubscript{3}-terminated
slabs of Bi\textsubscript{4}Se\textsubscript{3}(111) is shown in
Figure~1. A surface state crossing protected by MS only is indicated
by the red circle and the blue circle indicates a crossing which is
\textquotedbl{}dually protected\textquotedbl{} \cite{Rauch} by both
MS and TRS. The TSS within the blue-circled region lie between the
BVB and HBCB. These TSS were the primary focus of previous investigations
\cite{Bi4Se3S_2012,QDG2013}. The TSS that cross within the red-circled
region, which are the focus of the present work, lie above the HBCB.
The parity invariants of the bulk band structure counted up to the
HBCB were previously determined to be +1, +1, +1, and \textminus{}1
at the $\Gamma$, Z, L, and F points of the BBZ, respectively
\cite{Bi4Se3S_2012}. The product of the parity invariants is \textminus{}1,
which characterizes the gap above the HBCB as a STI-type \cite{Fu2007_WSTIS,FuInv}.
Applying the methods of ref. \cite{TeoBiSb_2008} to our case, an
odd number of TSS pairs are expected to exist between the surface
projections of F ($\bar{M}$) and $\Gamma$ ($\bar{\Gamma}$).
Indeed, we observe a single pair of TSS that cross each other along
the $\bar{\Gamma}\bar{M}$ azimuth and degenerate with different ends
of the bulk gap at $\bar{\Gamma}$ and $\bar{M}$. Group-theoretical
considerations indicate why this crossing is allowed even while the
surface states are seen to be gapped along the $\bar{\Gamma}\bar{K}$
azimuth.

At the (111) surface, the $R\bar{3}m$ symmetry of the crystal reduces
to C\textsubscript{3v}. For wave-vectors lying between $\bar{\Gamma}$
and $\bar{M}$, the point-group symmetry reduces to C\textsubscript{s},
which contains two irreducible representations characterized by mirror
eigenvalues of $\pm i$. Through the definition of the mirror operation
\cite{TeoBiSb_2008}, it is easily shown that the two irreducible
representations correspond to states of opposite spin-helicity, which
cannot hybridize with each other on the mirror-invariant $\bar{\Gamma}\bar{M}$
azimuth. This explains why the crossing circled in red is allowed
and, indeed, protected by the crystal's mirror symmetry. In contrast,
the point group of the wave-vectors along $\bar{\Gamma}\bar{K}$ is
C\textsubscript{1}. By symmetry, crossings between $\bar{\Gamma}$
and $\bar{K}$ are avoided, even for states of opposite spin-helicity.
The same is true for all wave-vectors in the SBZ which do not lie
on a mirror-invariant azimuth. The double-arrow in panel (a) indicates
what can therefore be understood as a hybridization gap $\Delta$
resulting from the avoided crossing of spin-helical surface states,
as will be discussed later. Note that the combination of C\textsubscript{3}
and time-reversal symmetry imply that the surface Fermi surface will
consist of six \textit{equivalent} pockets that each enclose a surface
state degeneracy point. This observation is consistent with the definition
of a STI material as put forward by Teo, Fu, and Kane \cite{TeoBiSb_2008}.

The $\overrightarrow{k}$-dependence of the gapped structure results
from competing SOC interactions which couple to different components
of the spin degree of freedom. This can be captured in a model for
the SOC Hamiltonian $H_{soc}\propto\left(\overrightarrow{p}\times\overrightarrow{\nabla}V\right)\cdot\overrightarrow{\sigma}$
using $k\cdot p$ theory. Polar coordinates are chosen with $\bar{\Gamma}$
as the origin and \textit{$\theta$} denoting the in-plane azimuthal
angle from the\textit{ k\textsubscript{x}}-axis, aligned to the $\bar{\Gamma}\bar{K}$
direction. If the energy-splitting of the TSS near $k=0$ is taken
to equal $2v\kappa$, then $H_{soc}$ to first-order in \textit{k}
couples the in-plane, tangential component of spin $\left\langle \sigma_{t}\right\rangle $
to the out-of-plane electrostatic potential gradient with a strength
$v$ as $H_{1}\left(\overrightarrow{k}\right)\equiv v(k-\kappa)\sigma_{t}$.
To third-order in \textit{k}, there appears a second term \cite{FuWarping_2009}
that couples the out-of-plane component of spin to the in-plane crystalline
potential gradient with a strength $\lambda$ as $H_{2}\left(\overrightarrow{k}\right)\equiv\lambda k^{3}cos(3\theta)\sigma_{z}$.
Together, 
\[
H_{soc}\left(\overrightarrow{k}\right)=\left[\begin{array}{cc}
\lambda k^{3}cos(3\theta) & iv(k-\kappa)e^{-i\theta}\\
-iv(k-\kappa)e^{i\theta} & -\lambda k^{3}cos(3\theta)
\end{array}\right]
\]
and we find the spin-orbit contribution to the dispersion 
\[
E_{soc}(\overrightarrow{k})_{\pm}\,=\,\pm\sqrt{v^{2}(k-\kappa)^{2}+\lambda^{2}k^{6}cos^{2}(3\theta)}
\]
where we refer to + and \textminus{} as the upper Dirac branch (UDB)
and lower Dirac Branch (LDB), respectively. At $k=\kappa$, the magnitude
of the gap between the DBs is determined solely by the second term
under the square root. The gap carries the sign of an \textit{f}-wave,
which changes at the mirror-invariant azimuths, signifying a change
in the \textit{z-}polarizations of the DBs (this could alternately
be described as a crossing of bands with positive and negative z-polarization).
The sign appears in the interband matrix element 
\[
\Delta=\left\langle -\left|H_{2}\left(\overrightarrow{k}\right)\right|+\right\rangle =-2\lambda k^{3}cos(3\theta)
\]
 for the spin-helical states typically invoked in the discussion of
simple, gapless TSS. In this sense, the crystalline anisotropy should
cause an interband SOC effect that gives rise to the gapped structure.
The model only differs from the TSS of the STI Bi\textsubscript{2}Te\textsubscript{3}
\cite{FuWarping_2009} in the location of the Dirac point. No extra
bands need to be inferred to achieve the complex, partially gapped
TSS described below.

\begin{figure}
\includegraphics[width=90mm]{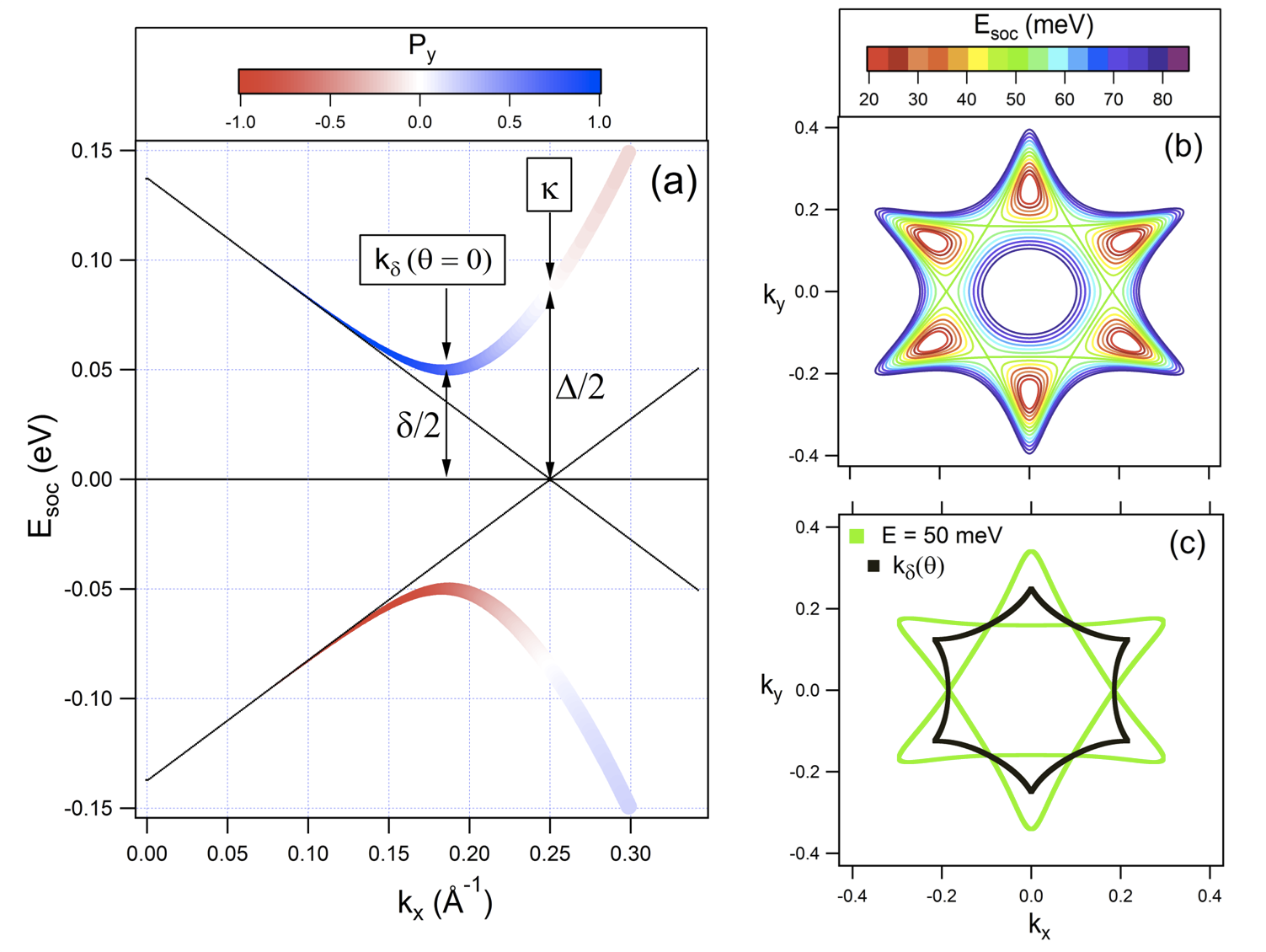}\caption{Electron-structure given by the model Hamiltonian for parameters stated
in the text: Band structure near the anti-crossing point along $\bar{\Gamma}\bar{K}$
($k_{y}=0$) direction (a). The color scale (a, inset) indicates the
spin-polarization in $\hat{y}$ direction, while marker size indicates
the degree of z-spin-polarization. Constant energy contours (CECs)
of the upper Dirac branch (b). The color scale (b, inset) indicates
the energy \textit{E} of each contour measured with respect to the
Dirac point. The momentum-space contour of the gap minimum overlayed
onto the CEC at the Lifshitz transition energy is shown in (c).}
\end{figure}

Fig. 2 displays electron-structure calculated from the model Hamiltonian,
for which we have chosen the values \textit{v = }5.5 eV{\AA}, $\lambda$
= 55 eV{\AA}\textsuperscript{3}, $\kappa$ = 0.25 {\AA}\textsuperscript{-1}
in rough approximation of the Bi\textsubscript{2}Se\textsubscript{3}-termination
electron-structure. The reader should note the many omissions from
this model such as spin-orbital entanglement (which will disallow
a pure spin-eigenstate character for the TSS and limit the magnitude
of spin-polarization) \cite{LouiesSOE}, and higher-order interaction
terms \cite{fifthorder}. Fig. 2(a) shows the spin-resolved band structure
along the $\theta=0$ azimuth, with the bands corresponding to the
case $\lambda$ = 0 plotted in black. The minimum energy gap $\delta$
is located inside of $k=\kappa$, and we find it is the case that
at \textit{$k_{\delta}\leq\kappa$} for all $\theta$, as shown in
panel (c). It is telling to inspect the in-plane helical spin-polarization,
indicated by color scale, relative to the magnitude of the out-of-plane
polarization, indicated by marker size. For $k<0.1$ {\AA}\textsuperscript{-1},
the spin is helical, but as the contribution of $H_{2}$ grows, the
deviation from linear dispersion is accompanied by increasing \textit{z-}polarization.
At $k_{\delta}$ the contribution
of $H_{2}$ overtakes that of $H_{1}$. At $k=\kappa$, the\textit{
z-}polarization is 100\% and then attenuates to $\sim$90\% for $k>\kappa$,
where the spin-helicity has reversed. The scenario of competing SOC
interactions takes place throughout the SBZ, resulting in unusual
constant energy contour (CEC) shapes and topologies, displayed in
panel (b). If the electron-dispersion were determined solely by the
SOC in this model, there would exist two Lifshitz points \cite{Lifshitz1960}
in the chemical potential located at $\mu=\pm E(k_{\delta},\theta=0)$.
At these points, the Fermi surface topology changes from six pockets
enclosing the TSS degeneracies to one electron-pocket and one hole-pocket
enclosing $\bar{\Gamma}$. Near a Lifshitz point, straight edges in
the CECs are centered on the $\bar{\Gamma}\bar{M}$ direction, opposite
of what would be expected for simple, gapless TSS on a (111) surface
\cite{FuWarping_2009}. This same pattern appears in CECs probed by
ARPES, described in Fig.3(c) below.

\begin{figure*}[t]
\includegraphics[width=42mm]{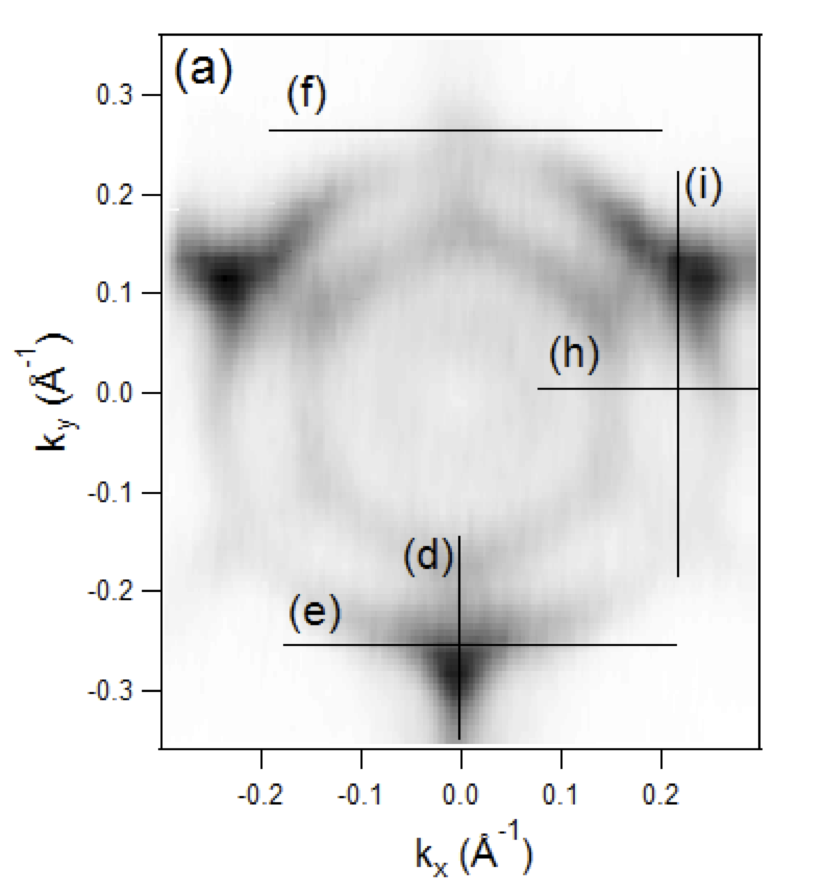}\includegraphics[width=75mm]{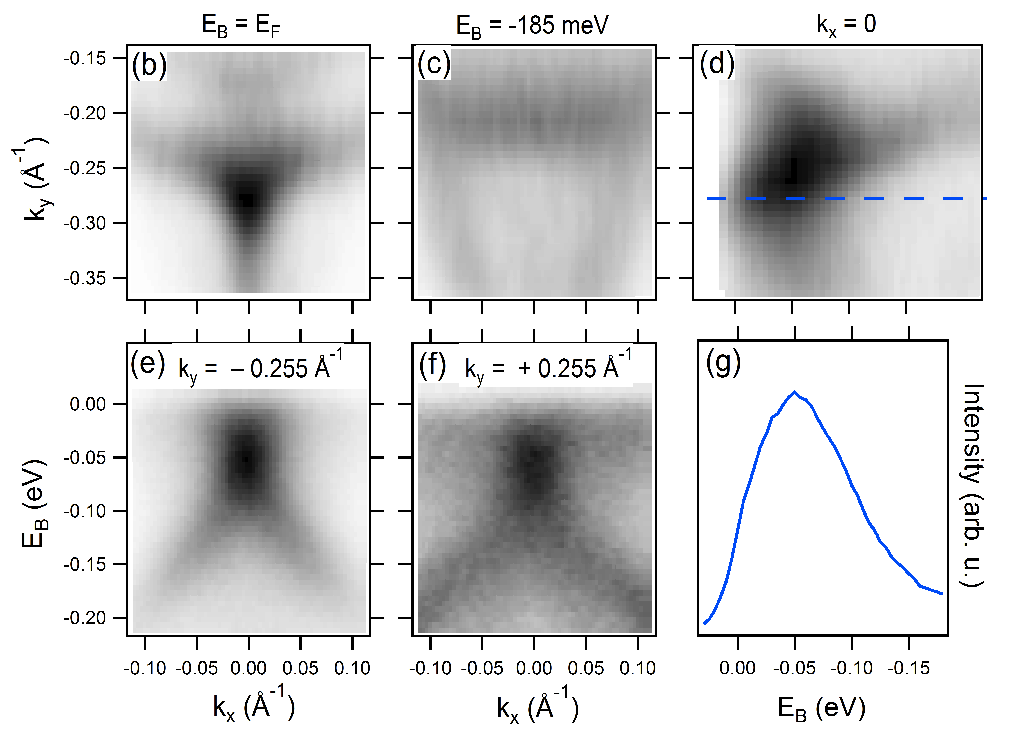}\includegraphics[width=65mm]{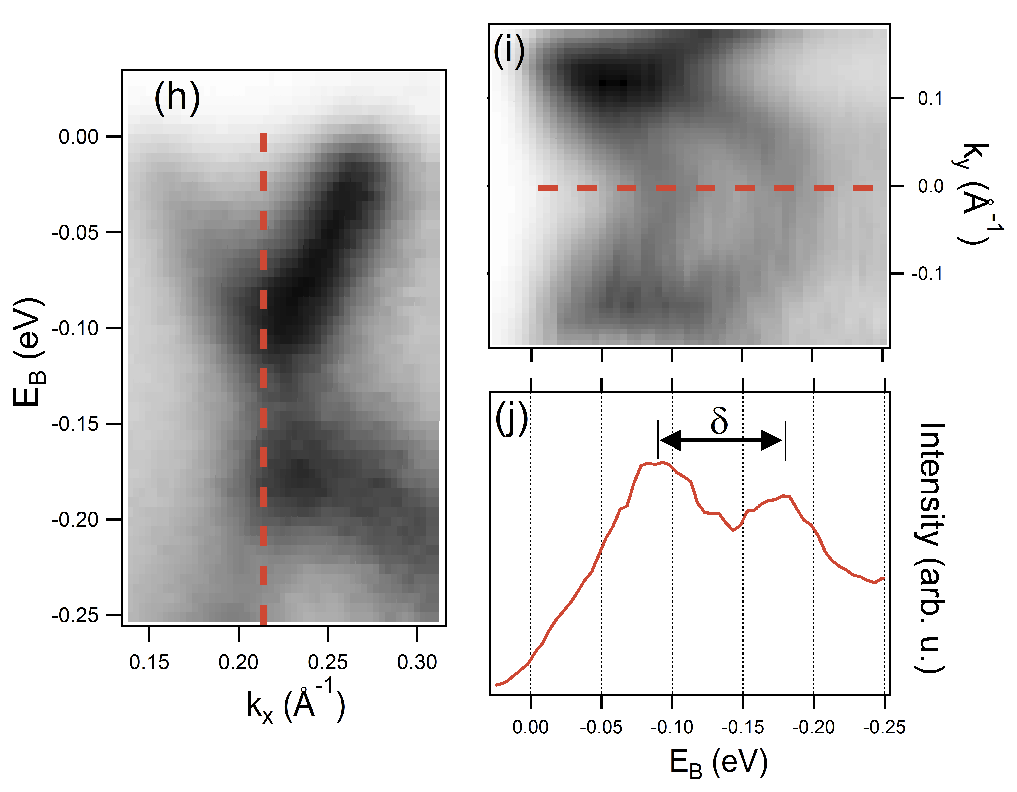}

\caption{ARPES electron-structure near the center of the 1\textsuperscript{st}
surface Brillouin zone: The Fermi surface (a,b), and constant energy
contour at -185 meV (c). Band structure parallel to $\bar{\Gamma}\bar{M}$
direction for $k_{x}=0$ (d) and $k_{x}=0.21$ {\AA}\textsuperscript{-1}(i).
Band structure parallel to $\bar{\Gamma}\bar{K}$ direction at $k_{y}=-0.255$
{\AA}\textsuperscript{-1} (e), $k_{y}=0.255$ {\AA}\textsuperscript{-1}
(f), and $k_{y}=0$ (h). Energy distribution curves at the degeneracy
point $(k_{x},k_{y})=(0,-0.255)$ {\AA}\textsuperscript{-1} (g) and the
saddle point of the lower Dirac branch $(k_{x},k_{y})=(0.21,0)$ {\AA}\textsuperscript{-1}
(j). The momentum-space locations for panels (d-e) and (h-i) are indicated
by lines overlayed onto the Fermi surface in (a). The gray-scale varies
from white-to-black following the minimum-to-maximum photoemission
intensity within each image individually.}
\end{figure*}

Fig.3 shows the ARPES spectra of the terraced surface of Bi\textsubscript{4}Se\textsubscript{3}(111)
collected using 70 eV photons. Previous photoemission electron microscopy
studies revealed that surface is terminated by Bi\textsubscript{2}
and Bi\textsubscript{2}Se\textsubscript{3} layers which are present
in approximately equal proportion \cite{QDG2013}. The Fermi surface
in panel (a) has three distinct branches enclosing $\bar{\Gamma}$.
It was previously determined that the circular, innermost branch is
derived from bulk conduction electrons and Bi\textsubscript{2}-surface
electrons, while the hexagonal branch outside of that is derived from
Bi\textsubscript{2}-surface electrons \cite{QDG2013}. The outermost
branch is derived from surface electrons of the Bi\textsubscript{2}Se\textsubscript{3}-termination,
which display a 3-fold enhancement of intensity due to ARPES matrix
element effects. The degeneration of TSS on the mirror-invariant $\bar{\Gamma}\bar{M}$
(\textit{k\textsubscript{\textit{x}}} = 0) line in the vicinity of
$k_{y}=\pm0.255$ {\AA}\textsuperscript{-1} at $E_{B}=-0.05$ eV is observed
in the band structure shown in panels (d-f). Crucially, the band structure
at $k_{y}=+0.255$ {\AA}\textsuperscript{-1} (e) and $k_{y}=-0.255$
{\AA}\textsuperscript{-1}(f) appears to be identical, confirming that
these bands consist of spin-polarized surface states, which have a
6-fold symmetry on the underlying 3-fold-symmetric lattice as required
by TRS. Comparing the Fermi surface in panel (b) to that of the -185
meV CEC (Fig.3(c)), we observe the formation of a \textquotedbl{}teardrop-shaped\textquotedbl{}
CEC that possess a straight edge centered on $\bar{\Gamma}\bar{M}$,
similar to what is predicted by the model Hamiltonian. Panels (h-j)
reveal a saddle-point in the LDB near $(k_{x},k_{y})=(0.21,0)$ {\AA}\textsuperscript{-1}
(indicated by red dashed-lines), where the minimum energy separation
$\delta$ between the DBs reaches a maximum value (with respect
to the in-plane azimuthal angle) of 85 meV. Comparing (h-i), it is
clear that the LDB reaches a minimum with respect to \textit{k\textsubscript{\textit{x}}
}and a maximum with respect to \textit{k\textsubscript{\textit{y}}}
at this point. This would yield a Van Hove singularity \cite{VanHove1953}
in the density of states. Interestingly, the UDB is at a local minimum
with respect to both variables at the same point in momentum-space.
The model Hamiltonian, which retains particle-hole symmetry, predicts
a saddle point in both DBs.

The observation of gapped surface states on a (111) surface with large
SOC is not without precedent, having been previously identified in
heterostructures with BiAg\textsubscript{2} surface alloys \cite{BiAgAu_2012},
however, the mechanism for the \textquotedbl{}interband SOC\textquotedbl{}
between the antiparallel, spin-helical surface states in that case
was left unspecified. The two-term model Hamiltonian approach shown
in this work could be extended to describe not only BiAg\textsubscript{2}
surface states, which are known to have a sizable coupling to the
in-plane crystal potential gradient \cite{firstBiAg}, but also many
other systems, whether topologically trivial or not, in which spin-helical
states intersect away from Kramer's momenta, such as Pb quantum wells
\cite{Slomski2013}. TIM, rather than topologically trivial materials,
may offer more robust platforms for studying this type of spin-gap
physics in 2D electron systems. What is lacking at this time is a
straightforward and reliable way of predicting if a given STI surface
will possess gapped TSS. The present results indicate a need to merge
different conceptualizations of the topological insulator (as defined
in the abstract sense band topology \cite{footnote}) in order to
determine the conditions necessary for this phenomenon.

Some have pointed out \cite{TeoBiSb_2008,Rauch,coaxialDirac} that
several well-known TIM possess bulk electron structure that can be
characterized as topologically non-trivial using separate methods.
The parity invariant method is used to characterize Z\textsubscript{2}
topological insulators \cite{Fu2007_WSTIS,FuInv}, which include STIs,
while mirror topological crystalline insulators (TCIs) are characterized
by a non-zero difference in the number of counterpropogating \textquotedbl{}edge
states\textquotedbl{} corresponding to a crystallographic mirror plane
\cite{TeoBiSb_2008,FuTCI,SnTetheory}. Teo, Fu, and Kane \cite{TeoBiSb_2008}
had considered that the crossing of TSS at non-TRIM was a possibility
for a strong Z\textsubscript{2} topological insulator surface, which
motivated them to develop the foundational theory for TCIs. This \textit{Letter}
has presented an experimentally realized case in which the concepts
of Z\textsubscript{2} topological insulators and TCIs have become
entangled beyond precedent; completely breaking the MS would allow
the TSS to become fully gapped, even while TRS remains unbroken. Surely,
the significance of the Z\textsubscript{2} topology in determining
the electronic physics at \textit{all }of the possible surfaces of
a STI should be revisited.
\begin{acknowledgments}
The financial support of the National Science foundation, Grants No.
NSF-DMR-0819860 and No. NSF-DMR-1104612, DARPA-SPAWAR Grant No. N6601-11-1-4110,
and the ARO MURI program, Grant No. W911NF-12-1-0461, is gratefully
acknowledged. The Advanced Light Source is supported by the US DOE,
Office of Basic Energy Sciences, under Contract No.DE-AC02-05CH11231.
Brookhaven National Laboratory is supported by the US Department of
Energy, Office of Basic Energy Sciences, Contract No. DE-AC02-98CH10886.\end{acknowledgments}


\begin{thebibliography}{10}
\bibitem{footnote}Any material with a band gap that remains open
at every point in the Brillouin zone can be regarded as an insulator
in the abstractions of topological band theory. The band structure
of a semimetal meeting this criterion can be \textquotedbl{}smoothly
deformed\textquotedbl{} into that of an insulator.
\end{thebibliography}
\end{document}